\documentclass[prl,showpacs,superscriptaddress,fleqn,preprint]{revtex4}

\usepackage{graphicx}
\usepackage{latexsym}
\usepackage{amsfonts}
\usepackage{amssymb}
\usepackage{amsmath}
\usepackage{bbm}

\newcommand{\be}{\begin{equation}}
\newcommand{\ee}{\end{equation}}
\newcommand{\bel}[1]{\begin{equation}\label{#1}}
\newcommand{\bea}{\begin{eqnarray}}
\newcommand{\eea}{\end{eqnarray}}
\newcommand{\bef}{\begin{figure}}
\newcommand{\enf}{\end{figure}}
\newcommand{\ba}{\begin{array}}
\newcommand{\ball}{\begin{array}{ll}}
\newcommand{\bacl}{\begin{array}{cl}}
\newcommand{\bacll}{\begin{array}{cll}}
\newcommand{\bal}{\begin{array}{l}}
\newcommand{\bac}{\begin{array}{c}}
\newcommand{\ea}{\end{array}}

\begin{document}

\title{Exact solution of the Bernoulli matching model of sequence alignment}
\author{V.B.~Priezzhev}
\affiliation{Laboratory of Theoretical Physics, Joint Institute for Nuclear
Research, 141980 Dubna, Russia}
\author{G.M.~Sch\"utz}
\affiliation{Institut f\"ur Festk\"orperforschung, Forschungszentrum J\"ulich, D-52425
J\"ulich, Germany }
\date{\today }

\begin{abstract}
Through a series of exact mappings we reinterpret the Bernoulli model of sequence alignment
in terms of the discrete-time totally asymmetric exclusion process with backward sequential
update and  step function initial condition. Using earlier results from the Bethe ansatz
we obtain analytically the exact distribution of the length of the longest common
subsequence of two sequences of finite lengths $X,Y$. Asymptotic analysis
adapted from random matrix theory allows us to derive the thermodynamic limit directly
from the finite-size result.
\end{abstract}
\maketitle

\section{Introduction}

Sequence alignment deals with the problem of identifying similarities between two different
sequences of objects, represented by "letters" from some "alphabet". This problem has
a long history in combinatorics and in probability theory where one wishes to
find the longest common  subsequence (LCS for short) between two random sequences
of letters \cite{combi}.
More recently, sequence alignment has become a central notion in evolutionary biology where
it is used to probe functional, structural
or evolutionary relationships between DNA or RNA strands or proteins \cite{bioinfo}.
In this setting one wishes to quantify how ``close''  two sequences of genetic information
are by identifying the LCS of the same gene in different species.

Given a pair of fixed sequences of $c$ letters of lengths $X$ and $Y$, the length of
their LCS is defined by the recursion \cite{LCSalgorithm, BundHwa}
\begin{equation}
L_{X,Y}=\mbox{max}[L_{X-1,Y},L_{X,Y-1},L_{X-1,Y-1}+\eta_{X,Y}]
\label{recursion}
\end{equation}
with the boundary conditions $L_{i,0}=L_{0,j}=L_{0,0}=0$ for all $i,j \geq 0$.
The variable $\eta_{X,Y}$ is 1 if the letters at the positions $X$ and
$Y$ match each other, and 0 if they do not. If one ignores the
correlations between different $\eta_{X,Y}$, and takes them from the
bimodal distribution $F(\eta)=p\delta_{\eta,1}+(1-p)\delta_{\eta,0}$,
one gets the Bernoulli matching (BM) model of sequence alignment
\cite{Boutet}. To get a model closest to the original LCS problem, one has to put $p=1/c$.
In the thermodynamic limit of infinitely long sequences this problem has been
studied in some detail. With $X=xN$, $Y=yN$, Sepp\"al\"ainen derived rigorously
the law of large numbers limit. Asymptotically the quantity $L_{X,Y}/N$ is a random
variable converging a.s. to a function of $p,x,y$ which he computed explicitly \cite{Sepp97}.
Using an exact mapping to a directed polymer problem,
complemented with scaling arguments, it was shown more recently \cite{MN} that
asymptotically the quantity $L_{X,Y}$ is a random variable of the form
\begin{equation}
\label{LCSBMinf}
L_{X,Y} \stackrel{N\to\infty}{\longrightarrow} \gamma_p(x,y) N + \delta_p(x,y) N^{1/3} \chi
\end{equation}
where $\gamma_p(x,y), \delta_p(x,y)$ are known scale factors and
$\chi$ is a random variable drawn from the Tracy-Widom distribution
of the largest eigenvalue of GUE random matrices \cite{TW}.
In subsequent work \cite{MMN} some related quantities were obtained for the thermodynamic limit
using a mapping to a 5-vertex model and applying the Bethe ansatz.

In this paper we compute analytically the exact distribution of $L_{X,Y}$ for {\it finite} sequences
by a mapping of the BM problem to a stochastic exclusion process. The mapping of the sequence
alignment problem onto the asymmetric exclusion process has been proposed in \cite{Bund}.
The hopping dynamic considered in \cite{Bund} is the asymmetric exclusion process with
sublattice-parallel update which admits a transfer-matrix formulation and diagonalization of
the matrix for finite $X$ and $Y$. Since our interest lies in an analytical solution for
arbitrary $X$ and $Y$, we choose another mapping onto a discrete-time
fragmentation process which is equivalent
to a totally asymmetric simple exclusion process with backward sequential
update \cite{backward, Priezz}. This allows us to use earlier results obtained
directly from Bethe ansatz \cite{RS} for this stochastic lattice gas model.
Specifically, we will express the probability that the length
of the LCS is
at most $Q$ by the probability that the number of jumps of a selected particle in the
exclusion process up to time $Y$ is at least $X-Q$.
We also outline how (\ref{LCSBMinf}) arises in the thermodynamic limit from the result
for finite sequences.

\section{Mapping of the BM model to an exclusion process}

\begin{figure}
\begin{center}
\includegraphics[width=0.4\textwidth]{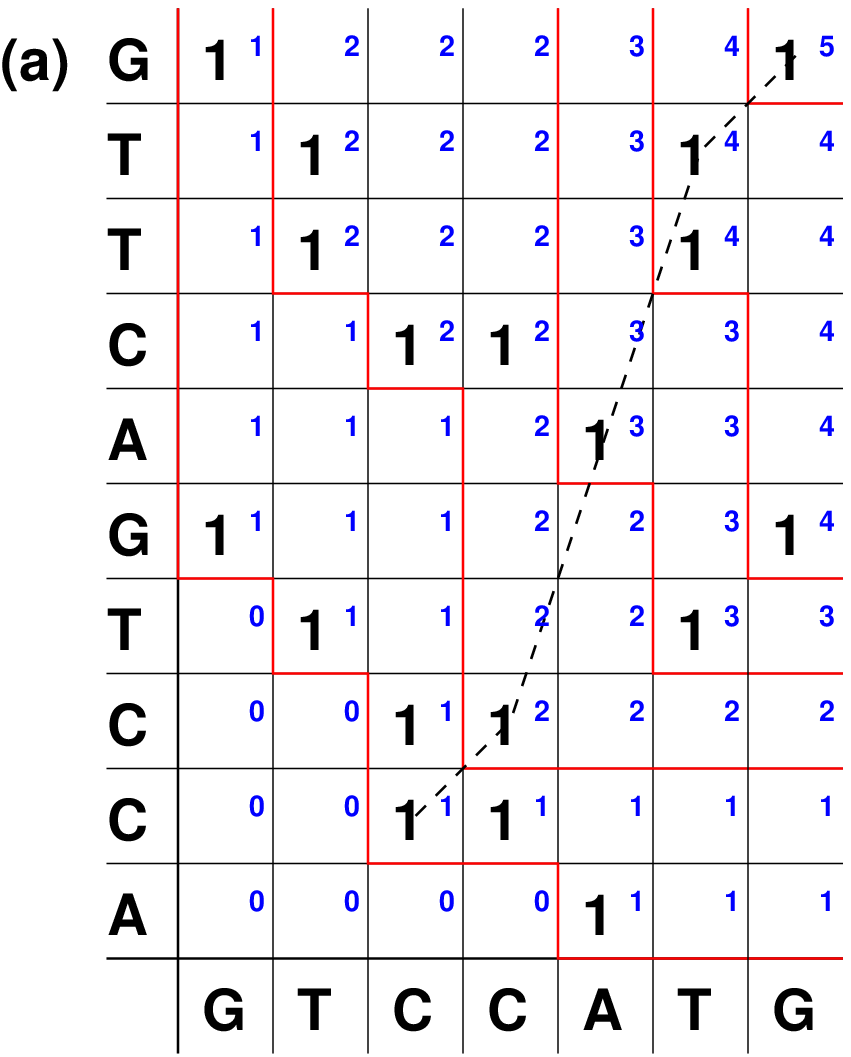}
\includegraphics[width=0.4\textwidth]{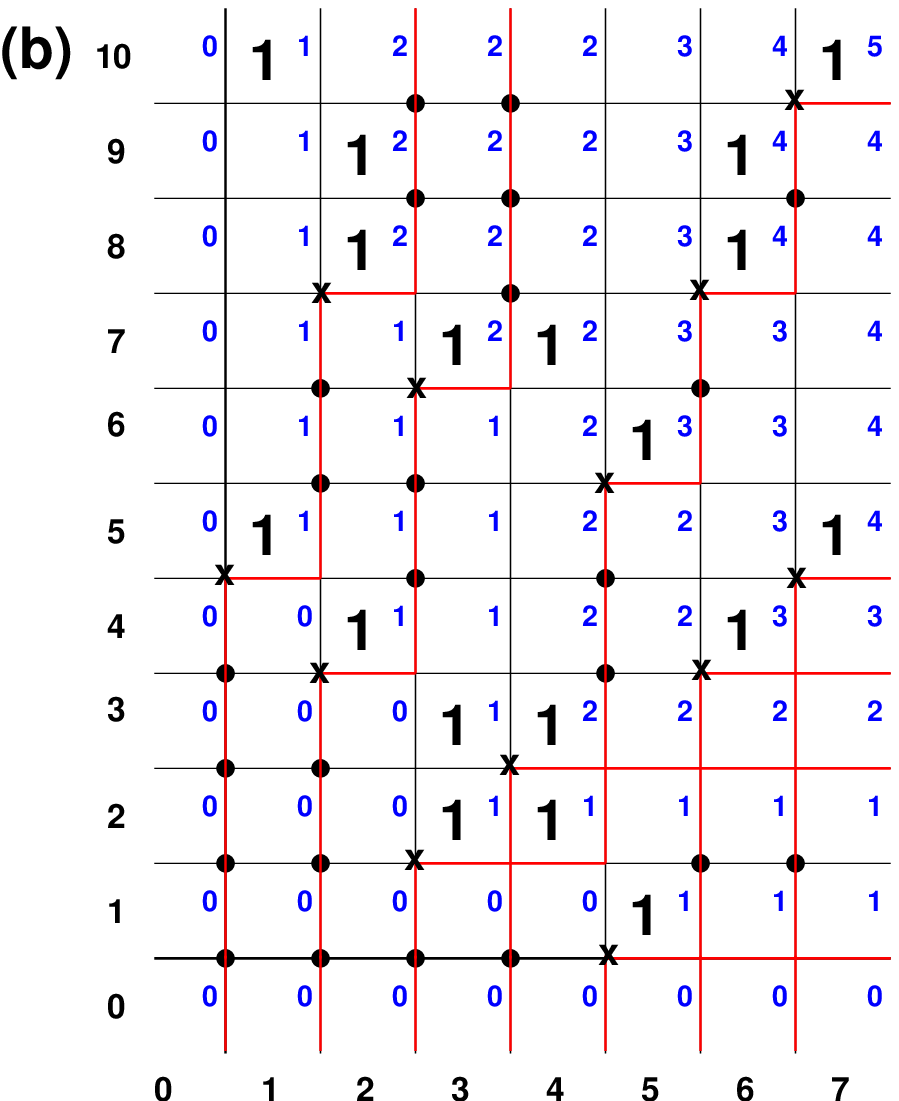}
\end{center}
\caption{
(a) Matrix representation of the LCS problem in sequence alignment for two sequences
of lengths $Y=10$ (vertical, bottom to top) and $X=7$ (horizontal, left to right).
Matches are denoted by a bold face 1 in the matrix. The small blue integers are the
solution of the recursion (\protect\ref{recursion}) with boundary conditions $L_{0,j}=L_{i,0}=0$
(not shown in (a), but in (b)). The red lines are the level lines that separate terraces of different height.
The dashed line follows the LCS $CCATG$ of length 5.
(b) Mapping to the five vertex model obtained by interchanging the colours of the vertical
lines and identifying lines with arrows as shown in Fig.~\protect\ref{fig2}(a). Vertex weights
$p$ are marked by an x, weights $1-p$ are marked by a bullet.
}
\label{fig1}
\end{figure}

In Fig.~\ref{fig1}(a) we illustrate the LCS problem in matrix form for two sequences of lengths
$X=7$ and $Y=10$ from
an alphabet of the four letters $A,C,G,T$  used in DNA sequencing. The vertical
sequence is read from bottom to top, the horizontal sequence is read from left to
right. Whenever two letters match there is a bold face 1 in the matrix.
The recursion (\ref{recursion}) (its solution is shown in small blue numbers in the matrix)
generates a terrace-like structure (red lines) where the number of terraces is
$L_{X,Y}$.
The boundary condition of the recursion amounts to assigning value 0 to the
boxes containing the letters of the two sequences and also to the empty lower
left starting corner.
Solving the recursion (\ref{recursion}), one can note that blue numbers in Fig.~\ref{fig1}(a)
appear with different statistical weights. Indeed, numbers at left corners of terraces appear
when $\eta_{X,Y}=1$ and have weight $p$. All the rest of numbers at edges of terraces
do not depend on $\eta_{X,Y}$ and have therefore weight $1$. All remaining numbers appear
when $\eta_{X,Y}=0$ having weight $(1-p)$.
It is useful to view the grid that defines the  matching matrix as a square lattice
with $X \times Y$ bulk sites, embedded in the rectangle  of size $(X+1) \times (Y+1)$.
Each square (defining the dual square lattice) is labelled $(i,j)$ with $0 \leq i \leq X$
and $0\leq j \leq Y$.
Due to the terraces, each site can take one of five different states. It may be
(i) traversed horizontally or vertically by a (red) terrace line
(ii) represent a left or right corner of a terrace, or (iii) be empty.

By construction, in the BM model each empty site has weight $q=1-p$,
each left corner of each line has weight $p$, and all remaining sites have weight 1.
This property allows for a mapping to a five-vertex model, see Fig.~\ref{fig1}(b).
For reasons that become apparent
below we define this mapping slightly differently from \cite{MMN} by interchanging
the colour of all vertical lines. The resulting pattern of intersecting black and red lines then
becomes isomorphic to the pattern of in- and outging arrows
in the five-vertex model with vertex weights given by the weights of the BM model.
One simply identifies black (red) horizontal lines with right-pointing (left-pointing)
arrows and black (red) vertical lines with up-pointing (down-pointing) arrows
as shown in Fig.~2(a).

\begin{figure}
\begin{center}
\includegraphics[width=0.4\textwidth]{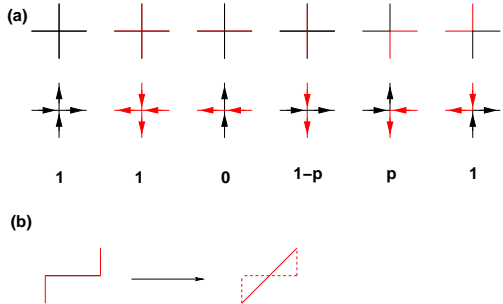}
\end{center}
\caption{\label{fig2}
(a) Mapping of line intersection to vertices of the six-vertex model which
is effectively a five-vertex model since one of the vertex weights is zero.
(b) A way to avoid line intersections in the five vertex model: if a
horizontal line has a left adjacent vertical line below and a right vertical
line above, it is replaced by the diagonal shortcut.}
\end{figure}

A similar terrace-like structure appears in the anisotropic 3D directed percolation
model (ADP) solved by Rajesh and Dhar \cite{Dhar}. A difference is that levels lines
in the ADP can overlap. The overlapping lines can be separated by successive shifts of
terraces, then one gets the five-vertex model with the same vertex weights as described above.
However, the shifts destroy the domain-wall boundary conditions which are essential for
thw exaxt solution of the BM model. Then, the analogy between the ADP and the BM models can be used
only in the thermodynamic limit as it was demonstrated in \cite{MN}.

It is useful to consider a further mapping of this 5-vertex model onto a
discrete-time stochastic process, considering the vertical direction as time and
horizontal one as discrete space.  To this end, we first turn the red vertex lines
(arrows pointing left or down)
into non-intersecting particle world lines by replacing a right-left turn with
a diagonal ``shortcut'', as shown in Fig.~\ref{fig2}(b). After a space reflection
$i \to i' = 1-i$ this yields a non-intersecting line ensemble
as shown in Fig.~3.

\begin{figure}
\begin{center}
\includegraphics[width=0.4\textwidth]{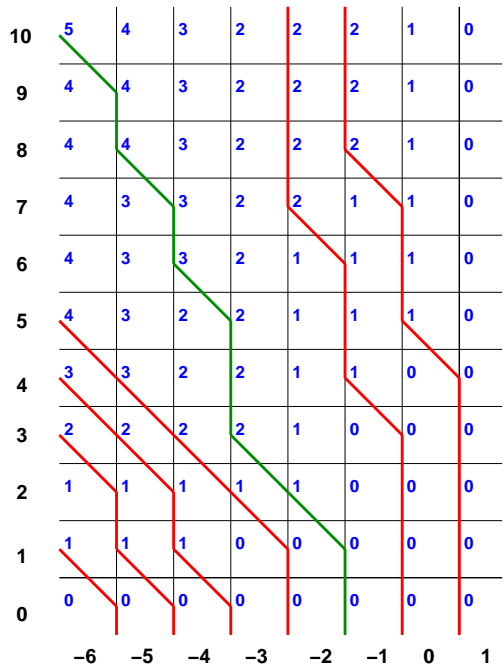}
\end{center}
\caption{\label{fig3}
Mapping of the five-vertex model to non-intersecting worldlines after space reflection
$i \to 1-i$.  The green worldline is the first line (seen from the right) which does not reach
the top of the grid.}
\end{figure}

\begin{figure}
\begin{center}
\includegraphics[width=0.7\textwidth]{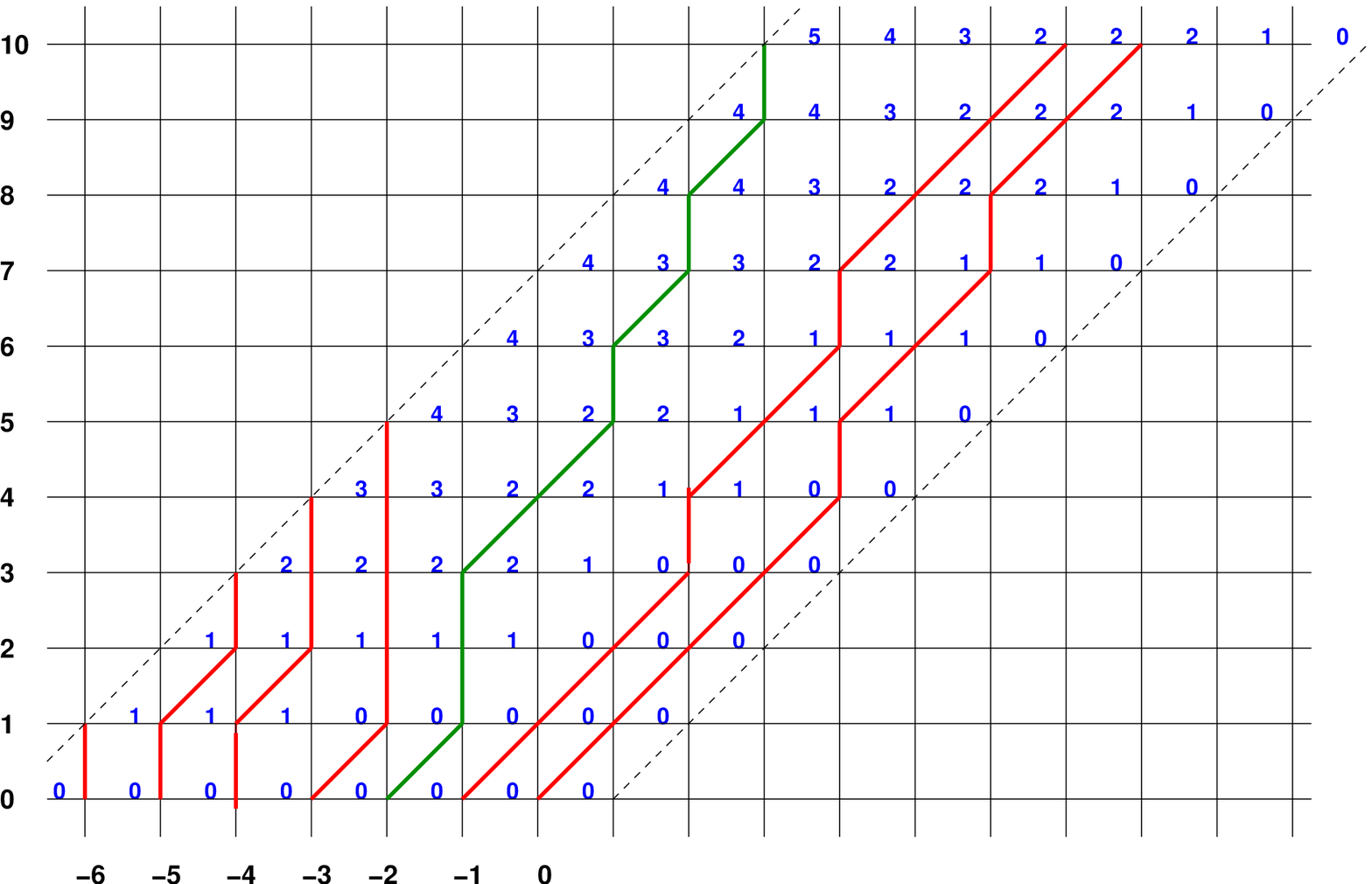}
\end{center}
\caption{\label{fig4}
Worldlines of the TASEP. The space-time grid of the
exclusion process is the square lattice $(i'+j+1/2,j)$ with new coordinates $(k,t)$
numbered by integers $k=i'+j$ and $t=j$. The green worldline is the
first line (seen from the right) which does not reach the target position $(k,t) = (Y-X+1,Y)=(4,10)$
that yields $L_{X,Y}$ for $X=7,Y=10$.}
\end{figure}

A final mapping is aimed to obtain the line ensemble of
particle world lines of the discrete-time totally asymmetric
exclusion process (TASEP) with the backward sequential update,
introduced in \cite{backward} and solved in \cite{Priezz}. To
this end, we consider each trajectory in Fig.~\ref{fig3} and replace each move upward by a diagonal move
right and each diagonal move left by a move upward (Fig.~\ref{fig4}). In a more formal way, we consider
a new square lattice $(i'+j+1/2,j)$ and draw new trajectories using the correspondence
$(i'+1/2,j) \rightarrow (i'+j+1/2,j)$, see Fig.~4. The sites of the new lattice are denoted
by coordinates $(k,t)$ numbered by integers $k=i'+j$ and $t=j$.
By construction the red lines move upward or diagonally and define the
world lines of exclusion particles which jump only to the right.
The vertex weights assign the appropriate
probability to each path ensemble.

The backward sequential dynamics encoded in the vertex weights
may be described as follows. In each time particle position are updated
sequentially from right to left, starting from the rightmost particle.
Each step of a particle by one lattice unit in positive direction has
probability $1-p$, provided the neighbouring target site is
empty. If the target site is occupied, the jump attempt is rejected
with probability 1.  No backward moves are allowed, making the
exclusion process totally asymmetric.
The horizontal boundary condition of the original sequence matching
problem maps into an initial condition
where at time $t=0$ particles occupy consecutive dual lattice
points $-X+1 \leq k \leq 0$.
Since the motion of a particle is not
influenced by any particles to its left, we may extend the lattice to minus
infinity. The vertical boundary condition is equivalent to extending the
lattice to plus infinity, such that at time $t=0$ all sites $k>0$ are vacant.
Thus one has a TASEP on an initially half-filled infinite lattice with
step initial state. However, only the first $X$ particles contribute to
the statistical properties of the BM model.

In the exclusion picture the terrace height has a simple probabilistic
interpretation. It counts the number of world lines that intersects
with a diagonal in the square lattice starting from the point $(k,t)=(-x,0)$
(the left dotted line in Fig.~\ref{fig4}.  Hence, at any given time step,
the terrace increases at each site from right to left by one unit,
unless a world line has been crossed when going from right to left.
Therefore the number $n$ of trajectories
ending at time $t=Y$ and the length $L_{XY}$ of the LCS of the BM model
on the rectangle
$(X+1) \times (Y+1)$ are related by $L_{X,Y}=X-n$.

\section{Probability distribution of the LCS for finite sequences}

Our aim is the evaluation of the probability distribution $\Lambda^{Q}_{XY} =
\mbox{Prob}[L_{X,Y}=Q]$ of the Bernoulli model.
Having the TASEP interpretation of the original model, we need to evaluate an
appropriate sum over end points of trajectories of particles. To do this, we select
the first trajectory (counted from the right) which does not end at time $Y$ in the
target range of the dual lattice given by the top row $(i,Y$ with $1\leq i \leq X$
(the green line in Fig.4). An important observation is that the sum of weights of
all trajectories ending at times $T_1,T_2,\dots T_k<Y$ (all lines to the left of
the green line) is 1 for the conservation of probabilities in the TASEP. Then,
the distribution $\Lambda^{Q}_{XY}$ is the sum over the probabilities of
all trajectories with end points right of the green line and over end points
of the green line itself.
Hence of all $X$ particles only the rightmost $n+1=X-Q+1$ particles are relevant
for the computation of $\Lambda^{Q}_{XY}$. The initial positions of these
particles are $k_1=-X+Q, k_2=k_1+1,k_3=k_2+1,\dots,k_{n+1}=0$.

Following the relation between terrace height $Q$ and particle trajectories
as discussed above, we may consider the final positions
$x_1,x_2,\dots,x_{n+1}$ at the moment of time $Y$. By the construction, we have
$Y-X+1 \leq x_2<x_3<\dots<x_{n+1}\leq Y$ and $x_1\leq Y-X$.
We first consider $Q=X$. In this case no particle has reached site $Y-X+1$.
In particular, this implies that the first particle (initially at site $0$) has not reached
site $Y-X+1$. The complementary probability for this event is the probability
$P(Y-X+1,1,Y)$ that the
first particle has jumped at least $Y-X+1$ times up to time $Y$.
Hence
\begin{equation}
\Lambda_{X,Y}^{X}=1-P(Y-X+1,1,Y).
\label{partfunction1}
\end{equation}

Now consider $Q<X$. Then $\Lambda^{Q}_{XY}$ is the joint probability that
after $Y$ time steps all rightmost $X-Q$
particles (located initially on $(-X+Q+1,\dots,0)$) have reached sites $\geq Y-X+1$
and the next particle (located initially on $-X+Q$) has not reached site $Y-X+1$.
This is equivalent to the joint probability that the particle originally at $-X+Q+1$
has jumped at least $Y-Q$ times and the particle originally at $-X+Q$ has jumped not more
than $Y-Q$ times. By construction of the process this joint probability may be expressed
as the statistical weight of all paths where the particle initially at $-X+Q+1$ jumps at least
$Y-Q$ times minus the statistical weight of all paths where the particle initially at $-X+Q$
jumps at least $Y-Q+1$ times.

We have come to a known problem of the TASEP statistics
\cite{Johansson,RS}.
Consider an infinite chain, the left half of which is initially occupied by
particles while the right half is empty. The problem is to find the probability
$P(M,N,t)$ that the $N$th particle (counted from the right) of the infinite cluster hops
at least $M$ times
up to time $t$. With this quantity, we obtain for the partition function
the expression
\begin{equation}
\Lambda_{X,Y}^{Q}= P(Y-Q,X-Q,Y)-P(Y-Q+1,X-Q+1,Y)
\label{partfunction}
\end{equation}
for every $Q<X$.
For $Q=X$, Eq.(\ref{partfunction}) reduces to (\ref{partfunction1}). We remark
that Eq.(\ref{partfunction1}) may be viewed as incorporated in Eq.(\ref{partfunction})
in agreement with the notion that the transition probability in
an exclusion process with no particles (second argument of $P$ for $X+Q$) is equal
to 1 (this is  the trivial transition
from the empty lattice to the empty lattice).

To derive Eq.(\ref{partfunction}) more formally,
let $P(x_1,x_2,\dots,x_{n+1};k_1,k_2,\dots,k_{n+1}|t)\equiv P({\bf x^{(n+1)}};
{\bf k^{(n+1)}}|t)$ be
the probability that $n+1$ particles located initially at
$k_1,k_2,\dots,k_{n+1}$
will be at $x_1,x_2,\dots,x_{n+1}$ at time $t$. Then, the partition function
$\Lambda^{Q}_{XY}$ can be written as
\begin{equation}
\Lambda_{X,Y}^{Q}=\sum_{x_{n+1}=Y-Q}^{\infty}\dots
\sum_{x_2=Y-X+1}^{x_3-1}\sum_{x_1\leq Y-X} P({\bf x^{(n+1)}};{\bf
 k^{(n+1)}}|Y)
\label{partition}
\end{equation}
Given the positions of $n$ particles at $x_2,\dots,x_{n+1}$, the sum of conditional
probabilities for the first particle to reach any position $x_1<x_2$ is
\begin{equation}
\sum_{x_1<x_2}P(x_1|x_2,\dots,x_{n+1})\equiv \sum_{x_1<x_2}P(x_1|{\bf
 x^{(n)}})=1
\label{norm}
\end{equation}
and
\begin{equation}
\sum_{x_1\leq Y-X}P(x_1|{\bf x^{(n)}})=1-\sum_{x_1=Y-X+1}^{x_2-1}P(x_1|{\bf
 x^{(n)}})
\label{rest}
\end{equation}

The notations in (\ref{norm}) and (\ref{rest}) mean that probabilities to find a particle
at $x_1$ with respect to positions of $n$ other particles, in contrast to probabilities
$P({\bf x};{\bf k}|t)$ conditioned with respect to the initial conditions.

Using (\ref{rest}), we get the partition function in the form
\begin{equation}
\Lambda_{X,Y}^{Q}=\sum_{x_{n+1}=Y-Q}^{\infty}\dots
\sum_{x_2=Y-X+1}^{x_3-1} P({\bf x^{(n)}};{\bf k^{(n)}})-
\sum_{x_{n+1}=Y-Q+1}^{\infty}\dots
\sum_{x_1=Y-X+1}^{x_2-1} P({\bf x^{(n+1)}};{\bf k^{(n+1)}})
\label{twoparts}
\end{equation}
where the time argument $Y$ is omitted.
The probability $P({\bf x};{\bf k}|t)$  for the continuous time TASEP has been
found in \cite{Schutz} and for the discrete-time
TASEP with the backward sequential update in \cite{Priezz}:
\begin{equation}
P({\bf x};{\bf k}|t)=\det |F_{i-j}(x_i-k_j;t)|
\label{detformula}
\end{equation}
where
\begin{equation}
F_{m}(n;t)=\frac{1}{2\pi
 i}\int_{|z|=1-0}dz(p+\frac{q}{z})^t(1-z)^{-m}z^{n-1}
\label{defff}
\end{equation}
In terms of $P(M,N,t)$, the first sum in (\ref{twoparts}) is
 $P(Y-Q,X-Q,Y)$
and the second one is  $P(Y-Q+1,X-Q+1,Y)$ which gives (\ref{partfunction}) again.

For the TASEP with parallel update $P(M,N,t)$ was computed by Johansson
\cite{Johansson} who used combinatorial methods of the theory of symmetric groups.
For the present model with backward sequential update the solution was obtained
by R\'akos and Sch{\"u}tz using the Bethe ansatz method \cite{RS}. For more general
initial conditions, this problem has been solved by Nagao and Sasamoto
\cite{Sasamoto}.
The expression for $P(M,N,t)$ obtained in \cite{RS} by evaluation of sums of
$P({\bf x};{\bf k}|t)$ reads
\begin{equation}
P(M,N,t)=\frac{q^{MN}(1-q)^{-MN+\frac{N(N+1)}{2}}}{\prod^N_{j=1}j!(M-j)!}
\sum_{t_1,t_2,\dots,t_N=0}^{t-1}\{\prod_{j=1}^{N}\prod_{k=0}^{M-N-1}(t_j-k)
(1-q)^{t_j}\}\prod_{i<j}(t_i-t_j)^2.
\label{jumpdistribution}
\end{equation}
where $q=1-p$ is the jump probability considered in \cite{RS}.
With (\ref{partfunction1}) and (\ref{partfunction}) this gives the
exact distribution of the LCS in the Bernoulli matching problem.
The cumulative distribution
\begin{equation}
\Xi_{X,Y}^Q = \mbox{Prob}[L_{X,Y} \leq Q] = \sum_{M=0}^{Q} \Lambda_{X,Y}^M
\end{equation}
takes the simple form
\begin{equation}
\Xi_{X,Y}^Q = P(Y-Q,X-Q,Y)
\label{cumulative}
\end{equation}
with the natural convention that $P(Y,0,Y)=1$.
In \cite{RS} it was shown explicitly that $P(Y-Q,X-Q,Y) = P(X-Q,Y-Q,X)$
which for the BM model is expected by symmetry. The result
(\ref{cumulative}) provides a simple relation between the cumulative
distribution of the length of the LCS in the BM model and the distribution
of the time-integrated current in the backward-sequential TASEP
for the step function initial condition.
The probability that the length of the LCS is at most $Q$ is given by
the probability that the number of jumps across bond $Y-X$ up to
time $Y$ is at least $X-Q$.

\section{Thermodynamic limit}

We now turn to a brief discussion how the asymptotic results (\ref{LCSBMinf})
of \cite{Sepp97} and \cite{MN}
follow from the cumulative distribution (\ref{cumulative}). To this end we need
the asymptotic properties of $P(M,N,t)$ for large arguments. For parallel update
Johansson has derived the asymptotics using results from the random matrix theory
 \cite{Johansson}. By a transformation proved in \cite{RS} this yields
the asymptotics of the distribution $P(M,N,t)$ for the discrete time TASEP
with the backward sequential update. One finds
\begin{equation}
\lim_{N\rightarrow \infty}P([\gamma N],N,N\omega(\gamma,q)+N^{1/3}\sigma(\gamma,q)\chi)
=F_{GUE}(\chi)
\label{asympt}
\end{equation}
with
\begin{equation}
\omega(\gamma,q)=\frac{(\sqrt{p}+\sqrt{\gamma})^2}{1-p}
\label{omega}
\end{equation}
and
\begin{equation}
\sigma(\gamma,q)=\frac{p^{1/6}}{\gamma^{1/6}}\frac{(\sqrt{p}+\sqrt{\gamma})^{2/3}
(1+\sqrt{p \gamma})^{2/3}}{1-p}
\label{sigma}
\end{equation}
The function $F_{GUE}(\chi)$ is the Tracy-Widom distribution of the Gaussian unitary ensemble
\cite{TW}.

The form of (\ref{asympt}) indicates that given $N$ and $\gamma$, the non-trivial
scaling regime in time
is given by the third argument. In the present case, $t=Y$ and $\gamma$ are fixed and
we search for the scaling regime of $Q$.
In our notations, $M=Y-Q$, $N=X-Q$, $t=Y$ and $\gamma = (Y-Q)/(X-Q)$.
From Eq.(\ref{asympt})
we have
\begin{equation}
Y \sim (X-Q)\omega(\gamma,q)+(X-Q)^{1/3}\sigma(\gamma,q)\chi
\label{Y}
\end{equation}
To find the asymptotics of $Q$ for large $X$ and $Y$, we represent it in the form
\begin{equation}
Q(X,Y,q)=Q_0(X,Y,q)+R(X,Y,q)\chi
\label{Q}
\end{equation}
Then, the leading term $Q_0(X,Y,q)$ can be found from the equation
\begin{equation}
Y=\frac{X-Q_0(X,Y,q)}{1-p}\left(\sqrt{p}+\frac{\sqrt{Y-Q_0(X,Y,q)}}
{\sqrt{X-Q_0(X,Y,q)}}\right)^2
\label{equat}
\end{equation}
which gives
\begin{equation}
Q_0(X,Y,q)=\frac{2\sqrt{pXY}-p(X+Y)}{1-p}
\label{Q_zero}
\end{equation}
which coincides with the expression found by Sepp\"al\"ainen \cite{Sepp97} using probabilistic
methods.

The substution of Eq.(\ref{Q}) with Eq.(\ref{Q_zero}) into Eq.(\ref{Y}) leads to the
equation for $R(X,Y,q)$ which can be resolved in the leading order of $\chi$.
In the first order, we substitute $Q=Q_0$ into the second term of the RHS of Eq.(\ref{Y})
to get
\begin{equation}
(X-Q_0)^{1/3}\sigma(\gamma_0,q)=\left(\frac{\sqrt{p}XY}{(\sqrt{X}-\sqrt{pY})
(\sqrt{Y}-\sqrt{pX})}\right)^{1/3}
\label{second}
\end{equation}
where
\begin{equation}
\gamma_0=\frac{Y-Q_0}{X-Q_0}
\label{gamma_zero}
\end{equation}
The coefficient $c_1$ at $R(X,Y)\chi$ in the expansion of the first term
of the RHS of Eq.(\ref{Y}) is
\begin{equation}
c_1=\frac{(1-p)\sqrt{XY}}{(\sqrt{X}-\sqrt{pY})(\sqrt{Y}-\sqrt{pX})}
\label{c_1}
\end{equation}
Then, $R(X,Y,q)$ is the ratio $(X-Q_0)^{1/3}\sigma(\gamma_0,q)/c_1$ and we obtain
\begin{equation}
R(X,Y,q)=\left(\frac{p}{XY}\right)^{1/6}
\frac{(\sqrt{X}-\sqrt{pY})^{2/3}(\sqrt{Y}-\sqrt{pX})^{2/3}}{1-p}
\label{R}
\end{equation}
Both results Eq.(\ref{Q_zero}) and Eq.(\ref{R}) coincide with corresponding expressions
obtained in \cite{MN} from a comparison with Johansson's result for the directed
polymer problem \cite{Johansson}.

\section{Conclusions}

We have considered the Bernoulli matching model of sequence alignment with the
aim of deriving the exact probability  that longest common subsequence of two
sequences of finite lengths $X,Y$
has length $N$.  By a series of mappings we have transformed the matching
problem to the time evolution of the totally asymmetric simple exclusion process
with backward sequential update in a step-function initial state. In this mapping the
computation of the probability distribution turns into the distribution of the
time-integrated current through a certain bond. This problem has been solved
by R\'akos and Sch\"utz \cite{RS} by Bethe ansatz methods. Thus the desired result for the
Bernoulli matching model for finite sequences has been obtained in explicit form
\eqref{cumulative} through some coordinate transformations from the Bethe ansatz.

In the thermodynamic limit we recover the earlier results of Majumdar and  Nechaev
\cite{MN} through an asymptotic analysis where we use the fact that in the thermodynamic
limit there is a scaling form of the current distribution found by Johansson
\cite{Johansson} which involves the distribution of the largest eigenvalue
of the GUE ensemble of random matrices. Adapting this result to the present setting requires
again some nontrivial
coordinate transformation. We find the occurrence of an eigenvalue distribution of random
matrices (which is valid also for finite sequence lengths, but for the Laguerre ensemble)
intriguing.

\section{Acknowledgements}
We thank S.Nechaev, S. Majumdar and K. Mallick for many useful discussions.
We acknowledge the support of the DFG. V.P. appreciates also the support of the RFBR
grant No. 06-01-00191a.
Part of this work was done while G.M.S. was the
Weston Visiting Professor at the
Weizmann Institute of Science.

\end{document}